\documentclass{acm_proc_article-sp}
\usepackage{graphicx}

\begin{document}

\conferenceinfo{}{Bloomberg Data for Good Exchange 2016, NY, USA}

\title{Intelligent Pothole Detection and Road Condition Assessment}

\numberofauthors{4}
\author{
\alignauthor
Umang Bhatt\\
       \affaddr{Carnegie Mellon University}\\
       \affaddr{Pittsburgh, PA}\\
       \email{umang@cmu.edu}
\alignauthor
Shouvik Mani\\
       \affaddr{Carnegie Mellon University}\\
       \affaddr{Pittsburgh, PAz}\\
       \email{shouvikm@cmu.edu}
\alignauthor 
Edgar Xi\\
       \affaddr{Carnegie Mellon University}\\
       \affaddr{Pittsburgh, PA}\\
       \email{edgarxi@cmu.edu}
\and
\alignauthor 
J. Zico Kolter\\
       \affaddr{Carnegie Mellon University}\\
       \affaddr{Pittsburgh, PA}\\
       \email{zkolter@cs.cmu.edu}
}

\maketitle
\begin{abstract}
Poor road conditions are a public nuisance, causing passenger discomfort, damage to vehicles, and accidents. In the U.S., road-related conditions are a factor in 22,000 of the 42,000 traffic fatalities each year$.^{1}$ Although we often complain about bad roads, we have no way to detect or report them at scale. To address this issue, we developed a system to detect potholes and assess road conditions in real-time. Our solution is a mobile application that captures data on a car's movement from gyroscope and accelerometer sensors in the phone. To assess roads using this sensor data, we trained SVM models to classify road conditions with 93\% accuracy and potholes with 92\% accuracy, beating the base rate for both problems. As the user drives, the models use the sensor data to classify whether the road is good or bad, and whether it contains potholes. Then, the classification results are used to create data-rich maps that illustrate road conditions across the city. Our system will empower civic officials to identify and repair damaged roads which inconvenience passengers and cause accidents.

This paper details our data science process for collecting training data on real roads, transforming noisy sensor data into useful signals, training and evaluating machine learning models, and deploying those models to production through a real-time classification app. It also highlights how cities can use our system to crowdsource data and deliver road repair resources to areas in need.
\end{abstract}

% A category with the (minimum) three required fields
%\category{H.4}{Information Systems Applications}{Miscellaneous}
%A category including the fourth, optional field follows...
%\category{D.2.8}{Software Engineering}{Metrics}[complexity measures, performance measures]

%\category{H.4}{Information Systems Applications}{Miscellaneous TODO}

%\terms{TODO}

%\keywords{TODO}

\section{Introduction}
Potholes and poor road conditions are a nuisance to society, causing discomfort to passengers, damage to vehicles, and accidents. We endure and complain about bad roads, yet have no way to detect or report them at scale. Meanwhile, civic authorities are not always aware of present road conditions, and road repairs happen only intermittently. 

Due to this inaction from both the consumers (the public) and caretakers (civic authorities) of road infrastructure, poor road conditions have become pervasive, leading to severe consequences. In the U.S., road-related conditions are a factor in 22,000 of the 42,000 traffic fatalities each year.\cite{halsey2009} Besides this tragic cost to human life, damage to vehicles from potholes costs Americans \$3 billion a year to fix.\cite{economist2016}

A key reason for poor road conditions is the information gap between the public, who travel on bad roads, and civic agencies, which are in charge of road repairs. To bridge this gap, we built a system that uses smartphone sensors to classify road conditions and potholes in real-time. This system leverages the public's road experience to inform civic authorities about roads that need repair.

In this paper, we also present a novel approach of using a combination of gyroscope and accelerometer sensors to provide insight into the condition of the road being traveled on. An accelerometer measures the linear acceleration in the X, Y, and Z directions, while the gyroscope measures the rate of rotation in each direction. Enumerating the linear and rotational movement of the phone (and the car) via these two sensors, we want to accomplish two central tasks:
\begin{enumerate}
\setlength{\itemindent}{.1in}
    \item Classify road conditions (good road/bad road)
    \item Detect potholes (pothole/non-pothole)
\end{enumerate}

By combining road classification data with insightful road condition maps, this intelligent system will help authorities direct repair resources to where they are most needed. This will improve road conditions and greatly benefit the public.

\subsection{Related Work}

Several other papers have demonstrated the use of smartphone accelerometer data to classify potholes and road conditions, but our approach differs from others in its inclusion of gyroscope data. Additionally, the deployment of our models to a real-time mobile app and the ability to produce road condition maps make our system more practical than others. These papers are outlined briefly below.

Mednis, et al demonstrate in their paper "Real time pothole detection using Android smartphones with accelerometers" that smartphones can be used to detect pothole events. Using a classification scheme that flags accelerometer activity that crosses a certain z-axis threshold, their algorithm detects potholes with true positive rates as high as 90\%.\cite{Mednis2011}

P Mohan. et al present Nericell, a fleet of smartphones using an aggregation server to assess road conditions, as well as a set of algorithms to reorient a disoriented smartphone accelerometer along a canonical set of axes.\cite{mohan2008}

Eriksson, Jakob, et al use a crowdsourced fleet of taxis called Pothole Patrol, gathering accelerometer and GPS data to identify potholes and road anomalies with a mis-identification rate of 0.2\%.\cite{eriksson}

%\newpage
\section{Methodology}
Before classifying potholes and road conditions, we had to collect a considerable amount of training data. We built a system for collecting and labeling this data via two separate iPhone apps. Then, we applied various transformations on the raw sensor data to get a better signal for classification.

\subsection{Specifications}
All of the data was collected on a 2007 Toyota Prius with approximately 100,000 miles. Both smartphones used for data collection were iPhone 6Ss. One iPhone was used for collecting sensor data while the other for recording potholes. An iPhone suction-cup mount was used to place the iPhone collecting sensor data on the center of the windshield. 

\subsection{Variable Definition and Controls}
For both of the problems, we needed to establish test groups and control for confounding variables. In the road condition classification problem, we reduced the varying degrees of road conditions to two extremes: good road and bad road. We did multiple drives on poor quality roads and on good roads. There was no pothole annotation done on these routes.
 
For the pothole detection problem, a major confounding variable was the route used for data collection. Different routes could have varying numbers and quality of potholes. In order to control for this and ensure reproducible results, we collected data on a single route. This route had a mix of pothole-free and pothole-filled stretches and ensured that we produced a balanced dataset. We traversed the route, shown in Figure 1, in only one direction.

\begin{figure}[h!]
\centering
\includegraphics[width = 8cm, height=5cm]{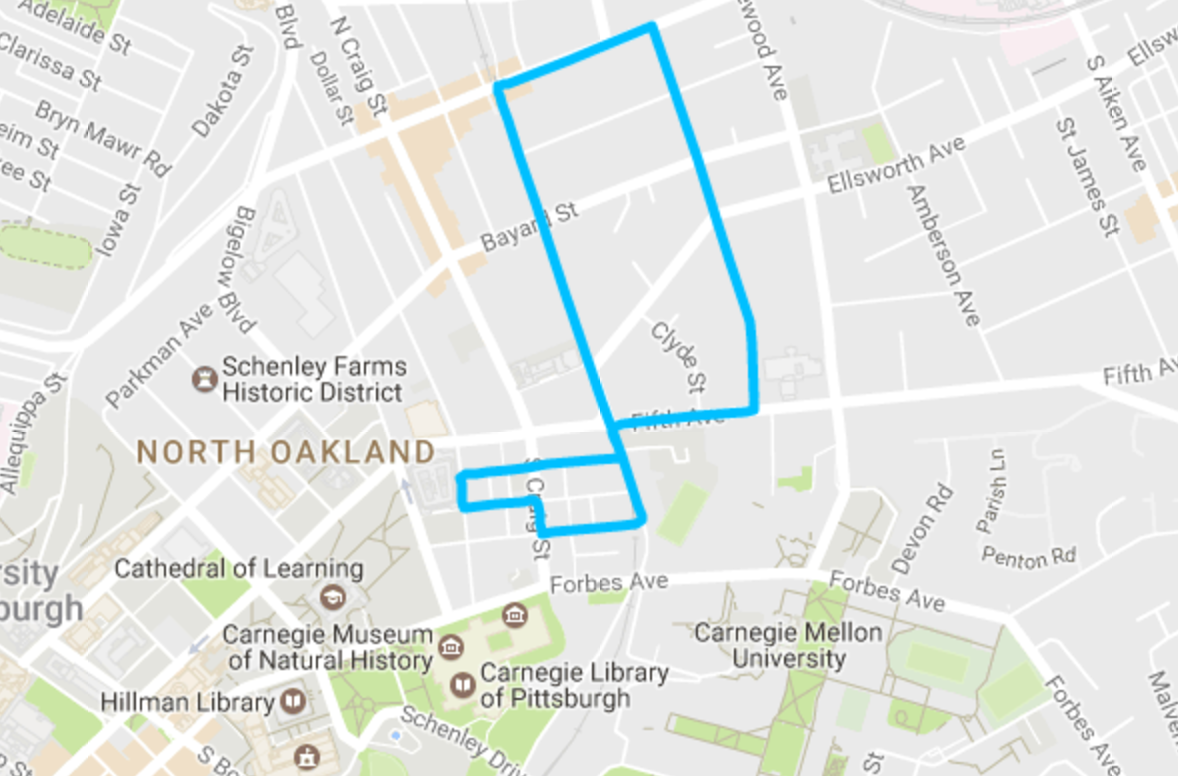}
\caption{Route for collecting training data}
\end{figure}

\subsection{Data Collection}
To facilitate the collection of training data, we built two iOS applications. One app collected sensor data (Figure 2). Specifically, five times per second, it recorded a UNIX timestamp, accelerometer data (x, y, z), gyroscope data (x, y, and z), location data (latitude and longitude), and speed. This app was run on an iPhone mounted near the center of the windshield of the car. It was used for both the good road/bad road and pothole detection problems, since both needed features on the car's movement.

\begin{figure}[h]
\fbox{\includegraphics[width = 4.5cm, height=8cm]{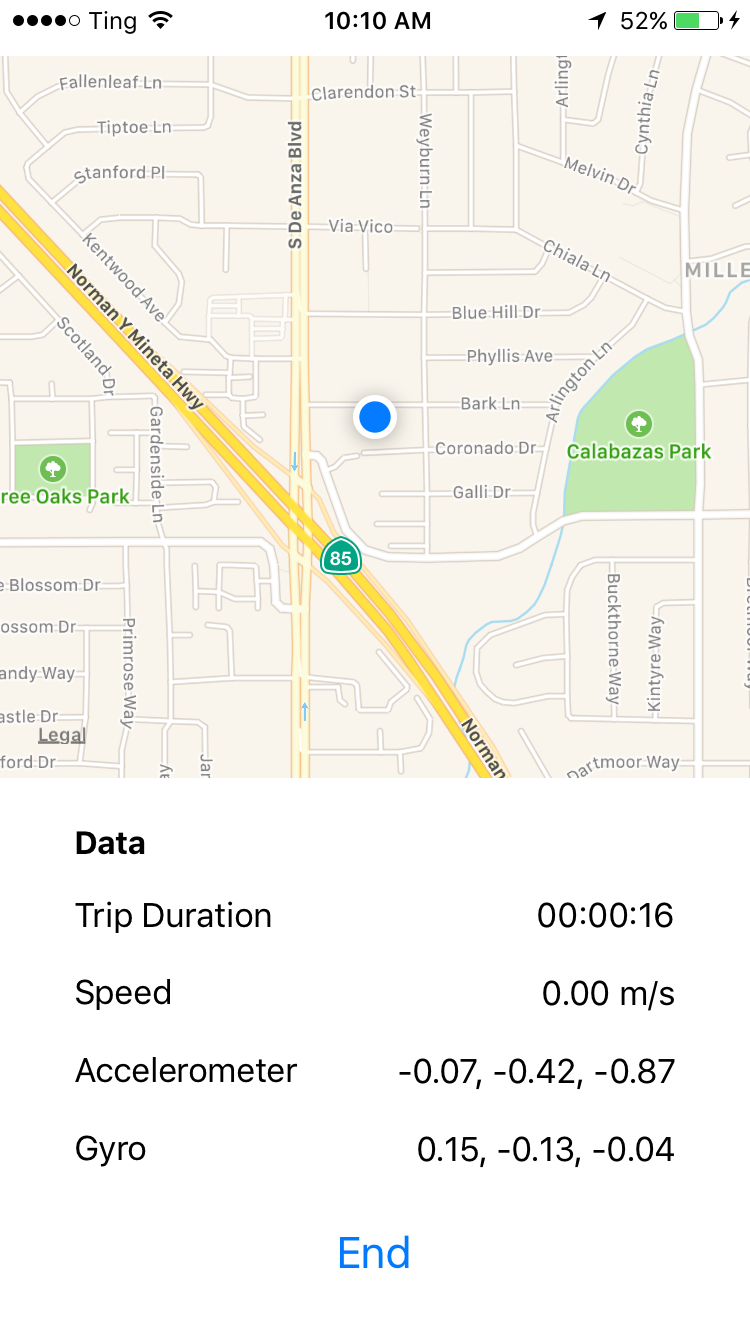}}
\caption{App 1 collected sensor data (timestamp, accelerometer, gyroscope, location, and speed)}
\end{figure}

\begin{figure}[h!]
\fbox{\includegraphics[width = 4.5cm, height=8cm]{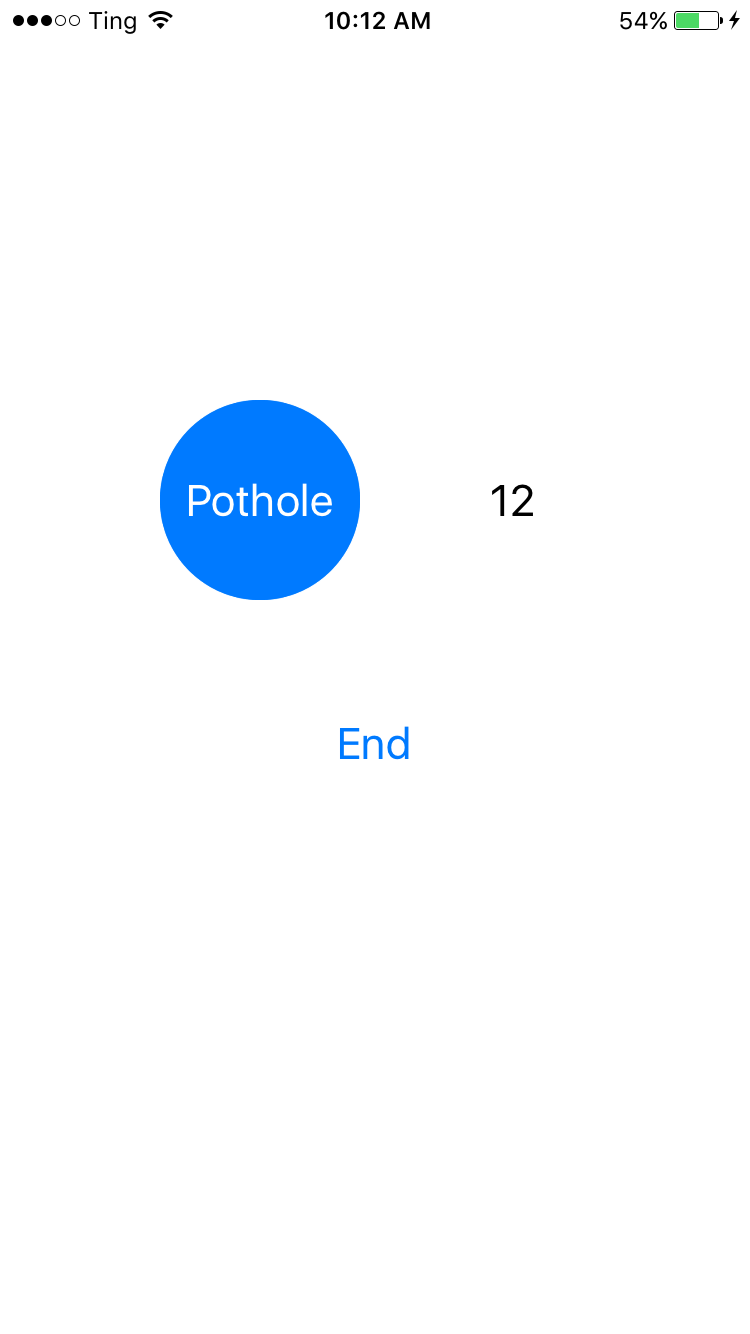}}
\caption{App 2 was used for labeling potholes and their timestamps}
\end{figure}

The second app (Figure 3) was used to annotate when a pothole was driven over - ideally, we wanted to get the exact time when a pothole was hit, but we will later discuss how we accounted for human error. This app was run on an iPhone given to a person on the passenger-side, whose job was to label the potholes. The passenger would simply click  a button when he or she felt a pothole, and the UNIX timestamp would be recorded. This app was used for the pothole detection problem and was run alongside the other iPhone collecting sensor data. 

To minimize undesired variance in our data collection, we set some controls. We used only one driver and one pothole recorder throughout the entire data collection process.

\subsection{Feature Engineering}
Once the individual training datasets (sensor data and pothole labels) were collected and combined, we had over 21,300 observations (71 minutes) of raw accelerometer and gyroscope readings as well as 96 labeled potholes. But since the sensor data was collected at a high frequency of 5 times per second, it was likely that the sensors captured some movements unrelated to vibrations caused by road conditions. So, the individual sensor data points were noisy and did not capture our variables of interest. 

To resolve this issue, we grouped data points into intervals and calculated aggregate features for each interval from the individual features. We created a set of 26 aggregate features for each interval which included: 

\begin{itemize}
\setlength{\itemindent}{.1in}
    \item{Mean accelerometer x, y, z}
    \item{Mean gyroscope x, y, z}
    \item{Mean speed}
    \item{Standard deviation accelerometer x, y, z}
    \item{Standard deviation gyroscope x, y, z}
    \item{Standard deviation speed}
    \item{Max accelerometer x, y, z}
    \item{Max gyroscope x, y, z}
    \item{Min accelerometer x, y, z}
    \item{Min gyroscope x, y, z}
\end{itemize}

Note: Aggregates for x, y, z dimensions for accelerometer and gyroscope sensors are three separate features.

For road condition classification, we decided to use an interval of 25 data points (5 seconds). We believed that 5 seconds was ample time to assess a small stretch of a road and classify it as good or bad. After creating the 5-second intervals and aggregate metrics, we attached the corresponding labels of good road (0) and bad road(1).

For pothole classification, we used an interval of 10 data points (2 seconds). Since potholes are sudden events, we hypothesized that a shorter interval would be able to capture them more accurately. For each interval, we attached the corresponding label of non-pothole (0) or pothole (1), depending on whether a pothole occurred during that interval.

Stitching together the sensor data and the labeled pothole data was a non-trivial problem, since labeling the potholes was itself an error-prone task. A person labeling potholes could be too late in clicking the pothole button or may click the button accidentally. By grouping the points into intervals, we addressed the former since a person could be slightly late in clicking the button but that interval would still be labeled as a "pothole" interval.

\section{Data Exploration}
We started by visualizing the data we gathered to see if we ourselves could notice any patterns. The goal of the following figures is to understand the data and come to meaningful conclusions that we can then transfer to our classifiers.

\subsection{Time Series}
Time series plots helped us understand whether there was a benefit in using the intervals and aggregate metrics instead of individual data points. Figure 4 shows the comparison between a good road and bad road using individual accelerometer readings (centered). Although there is clearly a difference between the plots, with the bad road plot having a higher variance, both of the datasets are noisy.

\begin{figure}[h!]
\centering
\includegraphics[width = 9cm, height=6cm]{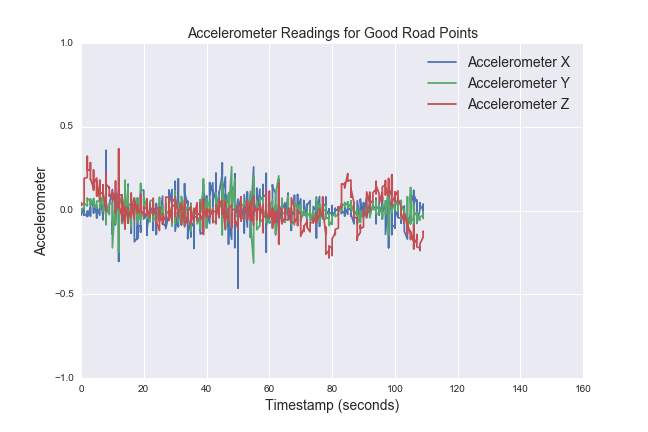}
\includegraphics[width = 9cm, height=6cm]{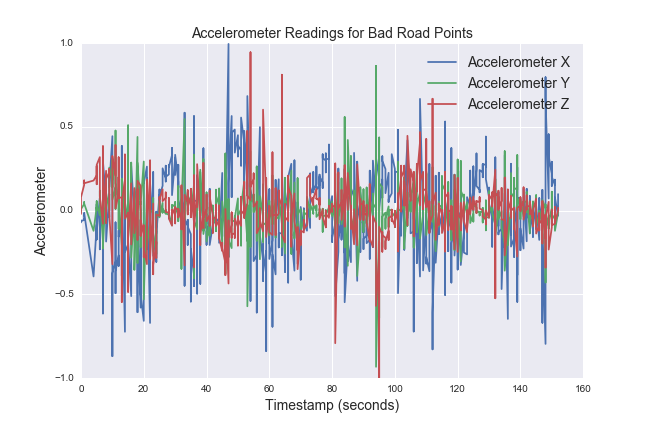}
\caption{Accelerometer readings (centered) for good vs. bad road}
\end{figure}

In contrast, Figure 5 shows the same data, but  grouped into intervals with their standard deviation accelerometer aggregates. Now, the difference between the good road and bad road data looks more pronounced. Doing this aggregation extracts the signal from the noise and produces a more stable set of features to use in our classifier.

\begin{figure}[h!]
\centering
\includegraphics[width = 9cm, height=6cm]{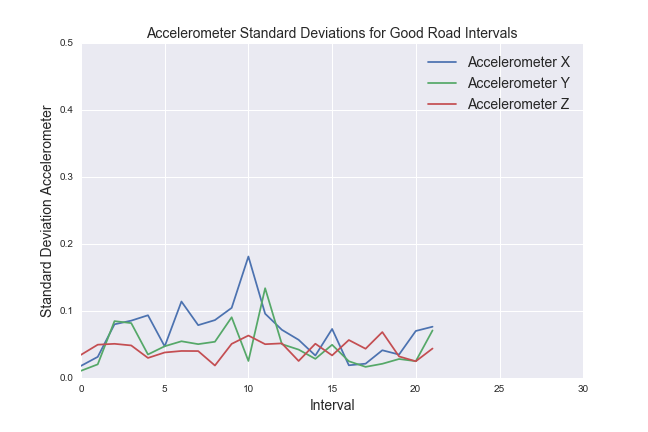}
\includegraphics[width = 9cm, height=6cm]{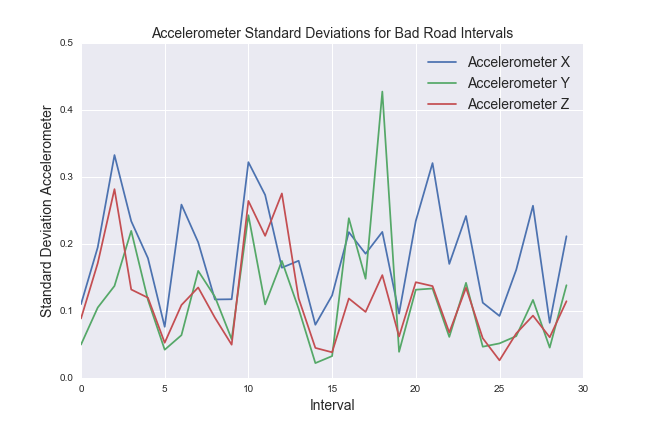}
\caption{Standard Deviation of accelerometer readings for good vs. bad roads}
\end{figure}

\subsection{Road Conditions Data Exploration}
Since we were not using time as a feature, we created 3D point clouds to visualize the data independent of time. By running principal component analysis (PCA), we reduced the 26-dimensional feature space of the intervals into three dimensions. Upon plotting the intervals and coloring them by their road condition label, we found a clear linear separation between good road and bad road intervals, as seen in Figure 6.

\begin{figure}[h]
\centering
\includegraphics[width = 9cm, height=6cm]{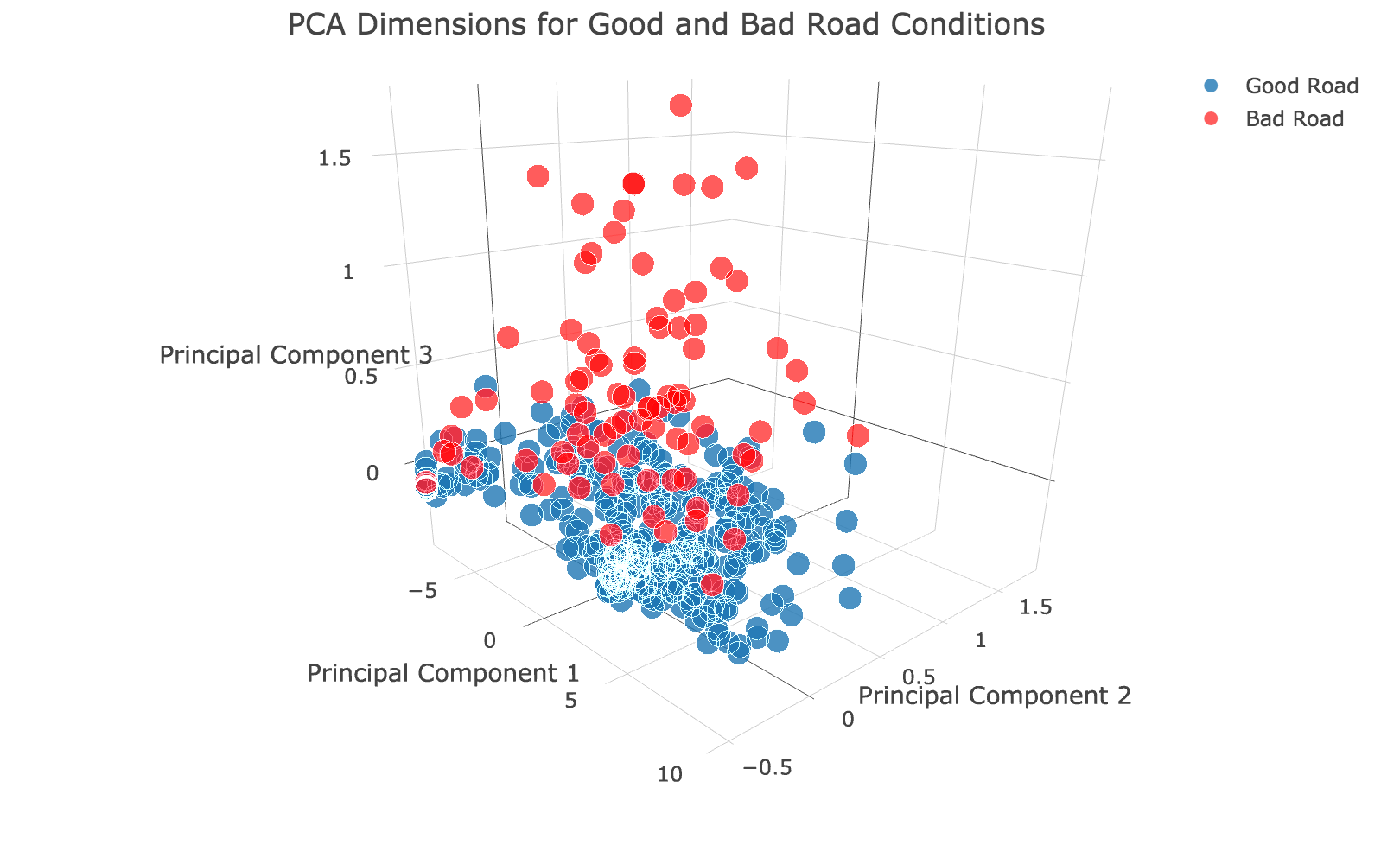}
\caption{PCA of features colored by road conditions}
\end{figure}

\subsection{Pothole Data Exploration}
Similarly, we ran PCA on the pothole data and plotted and colored the intervals in their reduced three dimensions. Once again, we observed a linear separation between the pothole and non-pothole intervals, as seen in Figure 7.

\begin{figure}[h]
\centering
\includegraphics[width = 9cm, height=6cm]{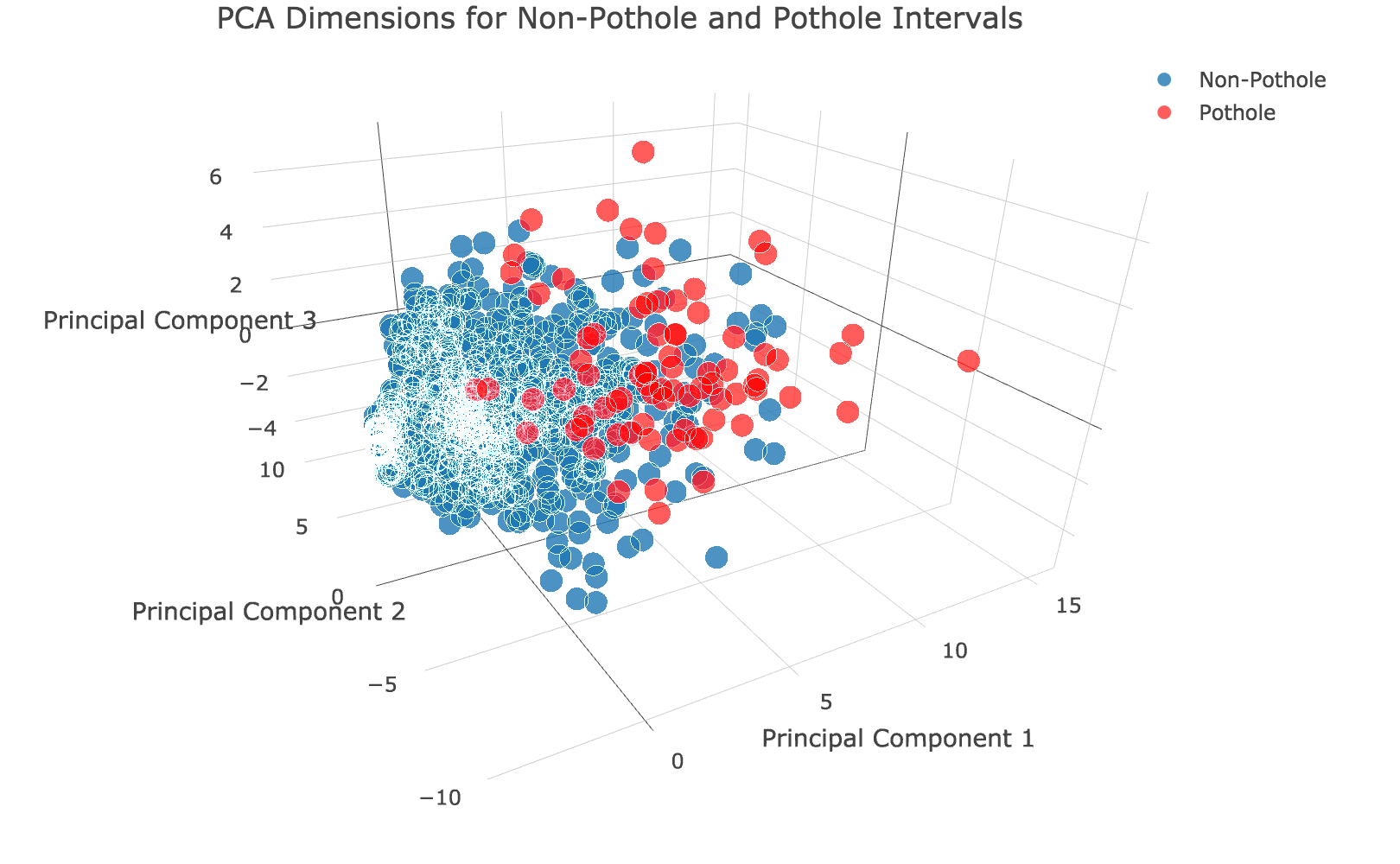}
\caption{PCA of features colored by presence of pothole}
\end{figure}

\section{Results and Analysis}
After combining data from all of our data collection trips and generating intervals and aggregate metrics, we trained several classifiers for both of the classification tasks.

\subsection{Road Condition Classification}
The road condition dataset was used to train and evaluate several classifiers, including support vector machine (SVM), logistic regression, random forest, and gradient boosting. The best results of each classifier can be found in the Appendix: Table 1. We tuned some of the parameters and hyperparameters for each classifier to get the best test set accuracy.

Overall, an SVM with a radial basis function (RBF) kernel and gradient boosting both achieved the highest test accuracy of 93.4\%. A baseline model that predicted "good road" for all instances would have achieved 82\% accuracy. SVM and gradient boosting beat the base rate in this problem

Figure 8 illustrates the selection of the regularization parameter C for the SVM classifier. Note that the training and test error are fairly close to each other at the chosen parameter value C=250, indicating that the model is performing well. However, more data and useful features could be helpful in further lowering this gap between the training and test error.

\begin{figure}[h!]
\centering
\includegraphics[width = 9cm, height=6cm]{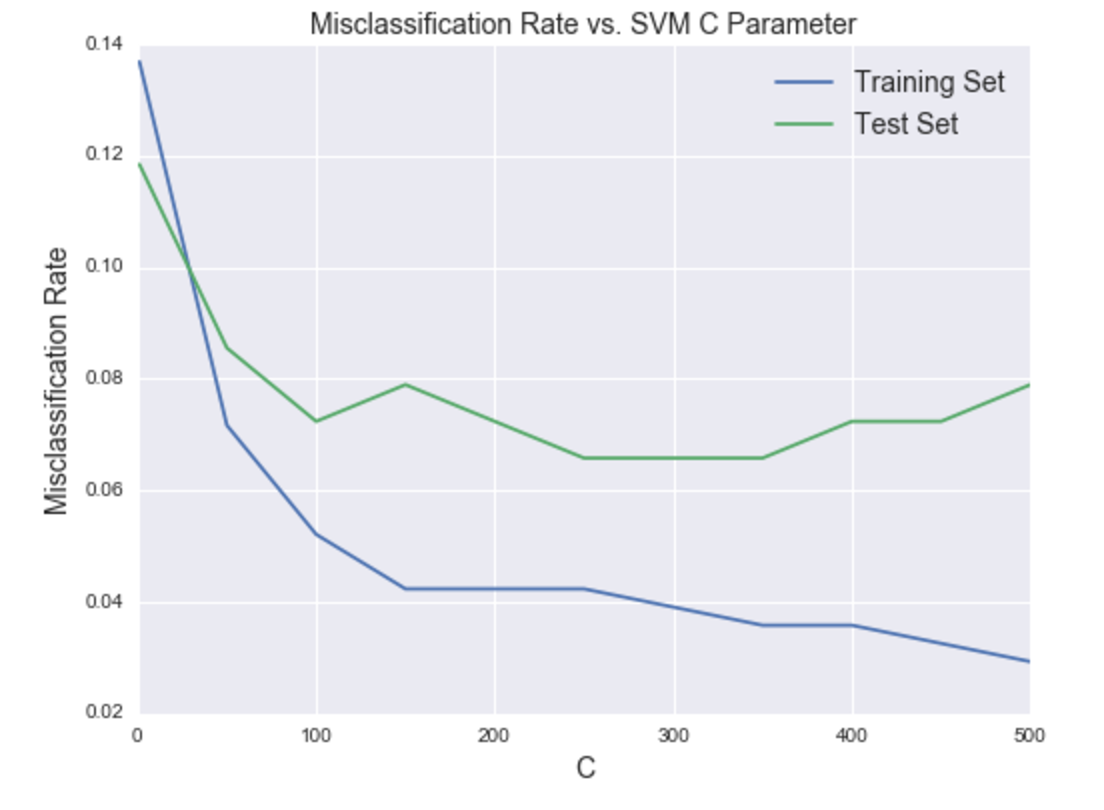}
\caption{Optimizing SVM regularization parmeter}
\end{figure}

\subsection{Pothole Classification}
Since potholes are rare events, there was a large class imbalance in our dataset. Even a naive model that always predicted "non-pothole" for a new interval would achieve 89.8\% accuracy on the classification task. So, in this problem, it was more important to optimize the precision-recall tradeoff than to focus on accuracy alone.

Like in road conditions classification, the best performing classifiers were SVM and gradient boosting, with accuracies of 92.9\% and 92.02\% respectively. These accuracies were somewhat better than the base rate of 89.8\% from the baseline classifier, so at least the classifiers were useful.

The SVM model performed the best in terms of accuracy, but we wanted to improve its precision-recall tradeoff. The precision-recall curve in Figure 9 illustrates all the combinations of precision and recall values for different thresholds on the SVM decision function. The red point in the figure represents the threshold we chose which gives us a precision of 0.78 and a recall of 0.42.

This choice is a good tradeoff between correctly flagging potholes (high precision) and detecting all true potholes (high recall). In this context, a precision of 0.78 means that when our model classifies an interval as having potholes, 78\% of those intervals actually have potholes. A recall of 0.42 means that our model correctly classifies 42\% of the true pothole intervals. Notably, the accuracy of the SVM model stayed at 92.9\% even though we changed the classification threshold to improve the precision-recall tradeoff.

\begin{figure}[h!]
\centering
\includegraphics[width = 9cm, height=6cm]{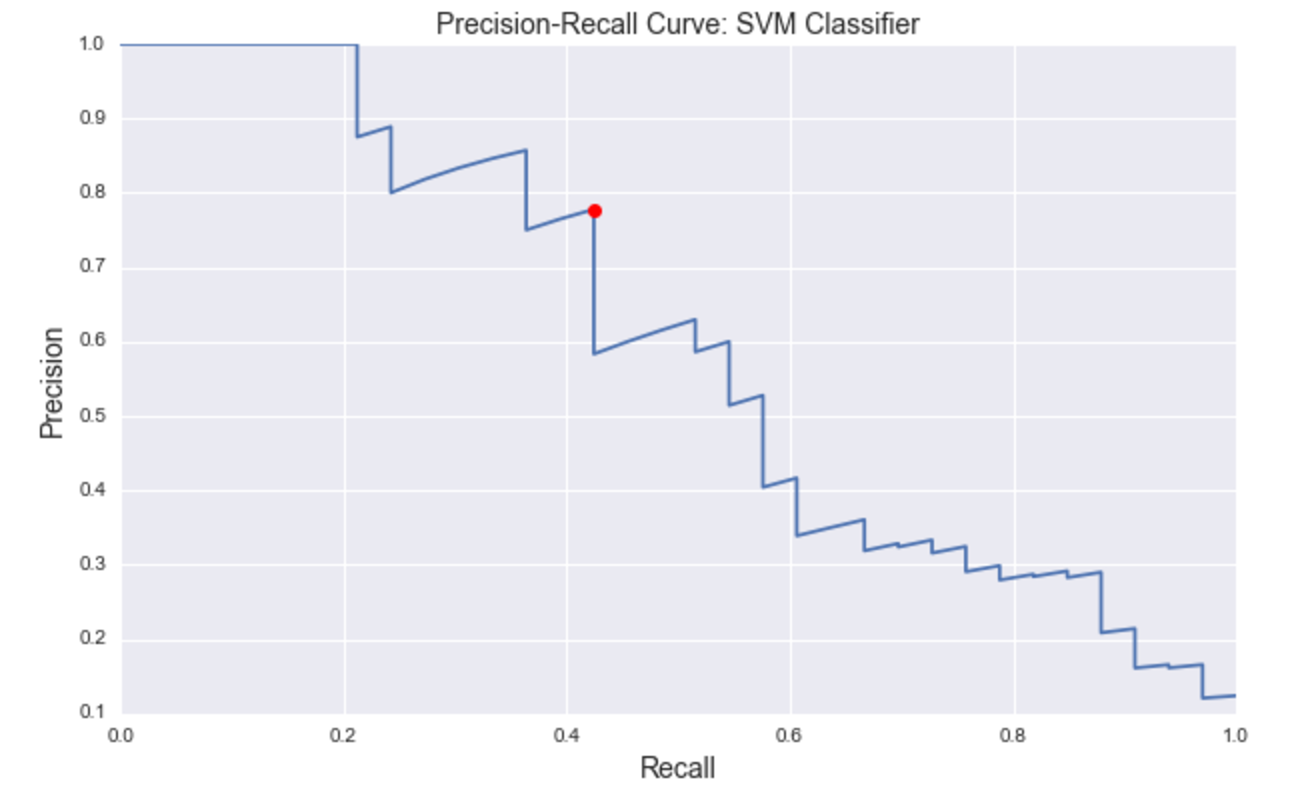}
\caption{Precision-recall curve for the SVM classifier. The red point is the precision-recall tradeoff we chose.}
\end{figure}

\section{Discussion}
In this report, we have presented a publicly available dataset and examples of using this data for road condition and pothole classification. This data and classification groundwork lays the foundation for real-time classification applications that have a high social impact. Additionally, it offers lessons for doing such work in the future as well as extension points to improve the work we have done.

\subsection{Real Time Classification Application}

After successfully building viable models for both road condition classification and pothole detection, we  developed a third iPhone app that does real-time classification for these tasks. While the previous apps were developed to collect training data, this app can be used to assess road conditions and detect potholes in the real world. The fitted SVM classifiers are deployed on a cloud-based web server, and the app is an interface for using the classifiers.

As the user drives, the iPhone app in Figure 10 collects data from the phone's sensors (accelerometer, gyroscope, latitude, longitude) and sends it to the classification server. The classification server applies the SVM models to classify the data and sends the results back to the phone app. The app then displays the classification results (good road/bad road, pothole/non-pothole) in 5-second intervals.

\begin{figure}[h!]
\fbox{\includegraphics[width = 4.5cm, height=8cm]{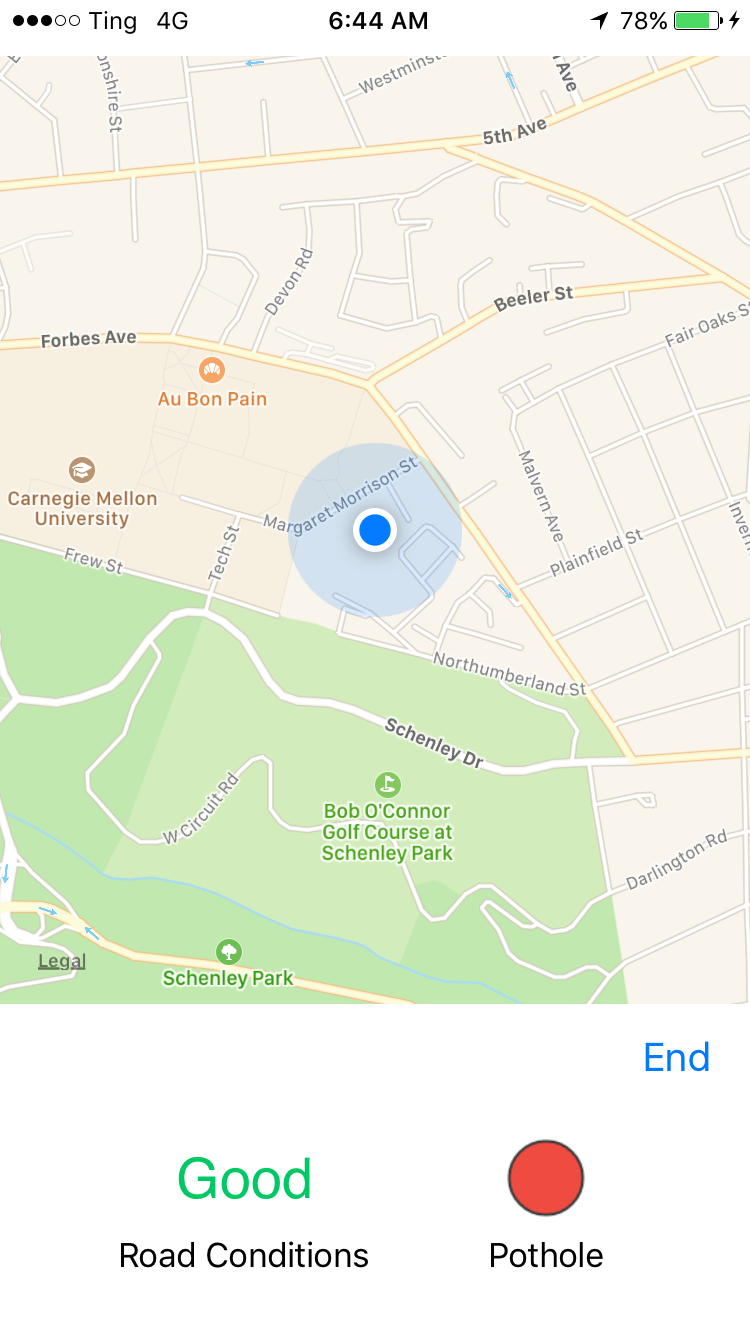}}
\caption{Real-time  app displays road condition and pothole classifications in 5-second intervals}
\end{figure}

\subsection{Road Condition Maps}

Using classification results from devices running the application, we can produce beautiful, data-rich maps of the city colored with potholes and road conditions, as shown in Figure 11.

\subsection{Social Good Application}

Crowdsourcing the classification and detection of road conditions and potholes could significantly improve the maintenance of road infrastructure in our cities. One could imagine the real-time classification app mentioned above being deployed to thousands of devices, constantly collecting road condition and pothole data from across a city. This data could be shared openly and combined with insightful road condition maps to help public works departments direct road maintenance resources to where they are most needed. 

According to Christoph Mertz, Chief Scientist of Roadbotics, smartphone sensors could also be put on public vehicles such as garbage trucks and post office vans, which cover the majority of a city's road network. The ability to create a less invasive method of detecting potholes and classifying road conditions would make it easier to disseminate the smartphone app, allowing for the creation of more detailed maps of a city's road conditions. Our work provides a basis for further work in crowdsourced public service. 

\begin{figure}[h!]
\centering
\includegraphics[width = 8cm, height=5cm]{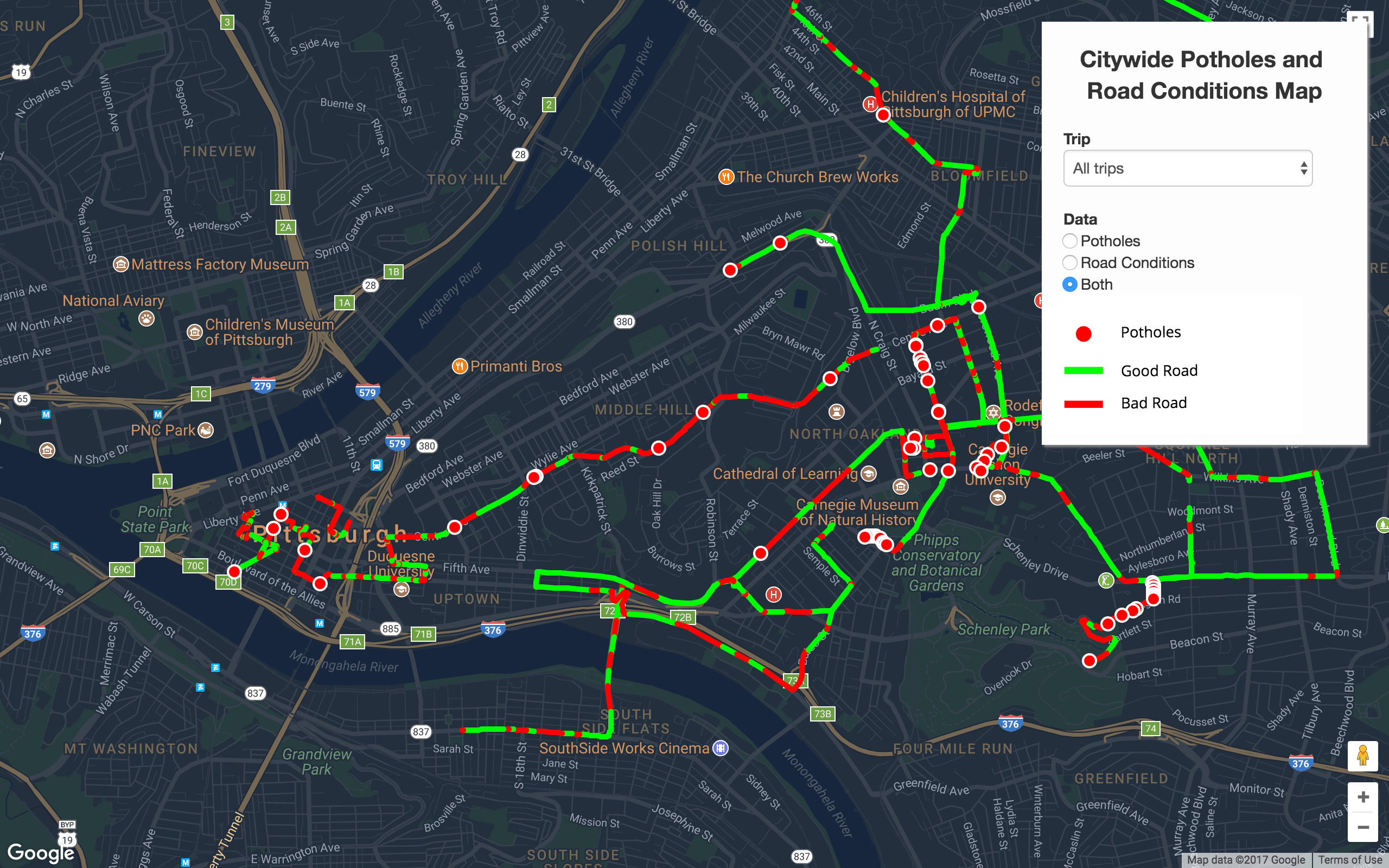}
\caption{Road condition map of Pittsburgh, PA showing classification results from the app}
\end{figure}

\subsection{Failures}
While performing this data exploration and analysis, we ran into many bumps (no pun intended) and had to pivot our approach and methodologies. Below is a collection of the failures we had to overcome in order to produce safe and sound results for this project.
\begin{itemize}
    \item From the inception of this project, we intended on building a classifier that works on all roads. Unfortunately, in order to build a reliable classifier, we would need sufficient data from roads of all types, which would take far longer than three months. Thus, we found it essential to select a specific route to classify on. Sticking to one route ensures that we collect enough data for a reliable classifier of that particular route. 
    \item Before tackling the precise classification of individual potholes, we found it helpful to understand road conditions. We wanted to answer the question: can we differentiate between a \textit{good} road and a \textit{bad} road? After proving the feasibility and the accuracy of road condition analysis, we felt comfortable and confident in moving to pothole classification.
    \item A major pivot point for us came when we divided our app into two separate apps. Instead of collecting sensor data and tagging potholes on the same phone, we had one app mounted onto the dashboard collecting data undisturbed, and the second app was given to the passenger who solely annotated when the car ran over a pothole.
    \item On multiple occasions, we lost our collected sensor data due to a lack of robustness in our original application, which needed to be loaded onto the phones once every two weeks. Had we done this project again, we would have invested time upfront into the applications to ensure they are reliable and robust during data collection.
    \item Since we are collecting sensor data five times per second, we ended up collecting over a thousand data points per time we traversed our route. When beginning our analysis on all our data from a given set of trips, we began to get bogus results; perplexed by what was happening under the hood, we quickly realized that trying to classify if a given fifth of a second occurs during a pothole is not insightful. Creating ten second intervals for detecting road condition and two second intervals for pothole detection proved to be more eye-opening. Over those intervals we were able to extract a multitude of features discussed above.
\end{itemize}

\section{Future Work}
There are many extension points from this initial data exploration and classification project. Below are a few examples of possible future work.
\begin{itemize}
    \item Expanding the route to collect training data on additional roads could only help by decreasing the variance of the models.
    \item Building a device to solely capture accelerometer and gyroscopic data  (though this does violate the aforementioned \textit{invasiveness}) would allow for less variance due to confounding variables.
    \item Working to control other confounding variables, like sudden changes in acceleration or mere braking, would make the features of the classifiers more robust.
    \item Calculating road condition scores (from 1-10, per se) would help extend this project beyond binary classification. These scores can then be mapped onto a given city/route to denote the conditions of roads comparative to other roads featured on the given map.
\end{itemize}

\section{Acknowledgments}
In addition to continual guidance and encouragement from Professor Zico Kolter of Carnegie Mellon University's School of Computer Science, we are also especially grateful to Professor Christoph Mertz for his insight to use gyroscope and accelerometer data to classify potholes. We would also like to thank Professor Roy Maxion, Professor Max G'sell, and Professor David O'Hallaron for their patient and thoughtful advice.

\nocite{*}
\bibliographystyle{abbrv}
\bibliography{main}

\end{document}